\documentclass[12pt,titlepage]{article}
\usepackage{bm, amssymb,pifont,cancel, amsmath,comment,color}
\usepackage{here}
\usepackage{cite}
\usepackage{graphicx}
\usepackage{subfigure}

\usepackage{appendix}
\usepackage{amsmath}
\usepackage{physics}
\usepackage{ dsfont }
\usepackage{mathtools}

\makeatletter

\@addtoreset{equation}{section}
\makeatother
 
\begin{document}

\begin{titlepage}
\null
\begin{flushright}
WU-HEP-20-02
\end{flushright}

\vskip 1cm
\begin{center}
\baselineskip 0.8cm
{\LARGE \bf
 {ALP Constraints in Gauged $\mathcal{N}=2$ Supergravity}}

\lineskip .75em
\vskip 1cm

\normalsize

{\large Shuntaro Aoki} $^{1}${\def\thefootnote{\fnsymbol{footnote}}\footnote[1]{E-mail address: shun-soccer@akane.waseda.jp}}, 
{\large } {\large Henry Liao} $^{1}${\def\thefootnote{\fnsymbol{footnote}}\footnote[2]{E-mail address: henryliao.physics@gmail.com}}

\vskip 1.0em

$^1${\small\it Department of Physics, Waseda University, \\ 
Tokyo 169-8555, Japan}

\vspace{12mm}

{\bf Abstract}\\[5mm]
{\parbox{13cm}{\hspace{5mm} \small

 We discuss a possibility of restricting parameters in $\mathcal{N}=2$ supergravity based on axion observations. We derive conditions that prepotential and gauge couplings should satisfy. Such conditions not only allow us to constrain the theory but also provide the lower bound of $\mathcal{N}=2\rightarrow\mathcal{N}=1$ breaking scale.

}}

\end{center}

\end{titlepage}

\tableofcontents
\vspace{35pt}
\hrule


\section{Introduction}\label{intro}

Extended ($\mathcal{N}\geq 2$) supergravity in four dimensions is well motivated by higher dimensional supergravity and string compactifications.
The aim of this paper is to investigate a possibility for constraining extended supergravity models from an observational data. In particular, we focus on the couplings between axion-like particle (ALP)\footnote{In this paper, the usage of ``axion'' and ALP are interchangeable.} and photon, which are intensively studied in the literature (see Ref.~\cite{Marsh:2015xka} for review). This kind of coupling usually exists in supergravity models if we regard neutral scalar fields in gauge kinetic function as ``ALPs", and a massless vector field as ``photon". 
In this paper, we show that some input parameters in extended supergravity can be constrained by the observational data on the ALP-photon coupling, by taking $\mathcal{N}=2$ supergravity as a toy example.

When we consider supersymmetry breaking in extended supergravity, several breaking scales appear in general. In the case with $\mathcal{N}=2$, there are two breaking scales, and in the following, we call the higher (lower) one as the first (second) supersymmetry breaking scale. The effects of the first supersymmetry breaking cannot be taken into account in the usual $\mathcal{N}=1$ supergravity description, where the first breaking scale is assumed to be high enough and its effect is negligible. However, if this is not the case, how the first supersymmetry breaking affects low energy physics would be a matter of central focus. For example, in Ref.~\cite{Abe:2019svc}, the effects of heavy modes arising from the first supersymmetry breaking is discussed, mainly focusing on gravitino decay process. In the following, we extend the analysis to the ALP-photon coupling and investigate the effects of the first suersymmetry breaking. Remarkably, we found that the first supersymmetry breaking scale enters in ALP couplings in a non-trivial way, which constrain the low scale first supersymmetry breaking models.

More specifically, we focus on a slightly generalized model of Ref.~\cite{Ferrara:1995xi}, where the first supersymmetry is broken but the second one remains unbroken (partial breaking). The model can of course include the effects of the first supersymmetry breaking (see also Refs.~\cite{Ferrara:1995gu,Fre:1996js,Itoyama:2006ef,Louis:2009xd,Louis:2010ui,Hansen:2013dda,Antoniadis:2018blk}) and easy to treat due to the unbroken $\mathcal{N}=1$ supersymmetry. For example, the stability of the vacuum is ensured. In addition, there are a lot of works which connect this model to the D-brane effective theory~\cite{Bagger:1996wp,Rocek:1997hi,GonzalezRey:1998kh,Tseytlin:1999dj,Burgess:2003hx,Antoniadis:2008uk,Kuzenko:2009ym,Ambrosetti:2009za,Kuzenko:2011ya,Ferrara:2014oka,Kuzenko:2015rfx,Ferrara:2016crd,Kuzenko:2017zla,Antoniadis:2017jsk,Farakos:2018aml,Cribiori:2018jjh,Antoniadis:2019gbd,Antoniadis:2019xwa}. Therefore, it would be appropriate 
for the first step of our research direction.

This paper is organized as follows. First, we introduce necessary elements in $\mathcal{N}=2$ supergravity and specify the model in Sec.~\ref{setup}. Next, in Sec.~\ref{spectra}, we derive mass spectra of bosonic sector. Then in Sec.~\ref{VScoupling}, ALP-photon coupling in our model is introduced, and we discuss the restrictions on the parameters and the supersymmetry breaking scale with given axion observations. All the aforementioned contents and remarks are summarized in Sec.~\ref{summary}. The technical details are collected in appendix. In Appendix~\ref{app:IntegratingOutTwoFormFields}, we derive the couplings between vector and scalar fields, and they are evaluated at assumed vacuum in Appendix~\ref{quadcub}.


\section{Setup} \label{setup}

In this section, we follow the representation and notation in Ref.~\cite{Andrianopoli:1996cm}, and introduce the setup in $\mathcal{N}=2$ supergravity. As mentioned in Introduction, we focus on the model~\cite{Ferrara:1995xi}, which consists of an Abelian vector multiplet, a hypermultiplet, and a gravitational multiplet. The bosonic components of the vector multiplet are a complex scalar $z$ and a four dimensional vector $ A_{\mu}$ with the spacetime index~$\mu=0,1,2,3$. Here we consider the multiple generalization of vector sector denoted by $i=1,2,\cdots,n_v$. The hypermultiplet contains four real scalars $b^u$, $(u=0,1,2,3)$, which parametrize quaternion-K\"ahler metric. Finally, the gravitational multiplet contains a graviton $g_{\mu\nu}$ and a graviphoton $A_{\mu}^0$. In the following sections, vectors will be denoted as $A_\mu^\Lambda$, $(\Lambda=0,1,\ldots,n_v)$, which combines vectors from vector and gravitational multiplets.


\subsection{Vector and hyper sector}
Here we only show the relevant ingredients of vector and hyper sector for later discussion (see Ref.~\cite{Andrianopoli:1996cm} for more detail). 

As for the vector sector,  the prepotential, $F$, which is a holomorphic and homogeneous
function of degree two with $n_v + 1$ complex variables $X^{\Lambda}$, acts as a master potential and takes form:
\begin{equation}
F=-i(X^0)^2f(X^i/X^0),
\end{equation}
where $f$ is an arbitrary holomorphic function in its argument which we take the following basis, 
\begin{equation}
X^0=1,\ \  X^i=z^i.  
\end{equation}
Here $z^i$ are identified as physical complex scalars in vector multiplets. With the prepotential defined, K\"ahler potential can be induced and takes expression given in
\begin{equation}
    \mathcal{K}=-{\rm{log}}\mathcal{K}_0,
\end{equation}
where $\mathcal{K}_0\equiv2(f+{\Bar{f}})-(z-{\Bar{z}})^i(f_i-{\Bar{f}}_i)$ and $f_i=\partial f/ \partial z^i$. Then  the K\"ahler metric is given as follows:
\begin{equation}
    g_{i{\bar{j}}} 
    \equiv \partial_{\bar{j}} \partial_i \mathcal{K}
    = \partial_i \mathcal{K} \partial_{\bar{j}} \mathcal{K} - \frac{1}{\mathcal{K}_0} (f_{ij}+{\bar{f}_{ij}}). 
\end{equation}
Besides, the gauge kinetic function $\mathcal{N}_{\Lambda \Sigma}$ is also determined by the prepotential and takes form:
\begin{align}
        \mathcal{N}_{\Lambda \Sigma}=
        \bar{F}_{\Lambda \Sigma} +
        \frac{2i {\rm{Im}}F_{\Lambda \Gamma} {\rm{Im}}F_{\Sigma \Pi} X^{\Pi} X^{\Gamma}}{{\rm{Im}}F_{\Gamma \Pi} X^{\Pi} X^{\Gamma}},
\end{align}
where $F_{\Lambda}=\partial F/\partial X^{\Lambda}$ and so on.

As for the hyper sector, we select the form of metric as shown in
\begin{equation}
    h_{uv} = \frac{1}{2(b^0)^2} \delta_{uv},
\end{equation}
which parametrizes a nonlinear sigma model on SO$(4,1)/{\rm{SO}}(4)$. Observe that it allows three commuting isometries:
\begin{equation}
    b^{\alpha} \rightarrow b^{\alpha} + c^{\alpha},\label{iso}
\end{equation}
where $\alpha=1,2,3$ and $c^{\alpha} \in \mathbb{R}$. Now, we 
consider to gauge this symmetry by $A_{\mu}^{\Lambda}$ as in Ref.~\cite{Ferrara:1995xi}, which will be reviewed in Sec.~\ref{sec:gauging}. 


\subsection{Gauging and relevant Lagrangian}\label{sec:gauging}

Let us discuss the gauging of Eq.~$\eqref{iso}$. To do so, we employ the embedding tensor formalism~\cite{deWit:2002vt,deWit:2005ub,Samtleben:2008pe,Trigiante:2016mnt}, which is useful to discuss a wider class of gauging. In this formalism, we formally introduce a double copy of the vector fields. In our system, we already have $n_v+1$ vector fields~$A_\mu^\Lambda$, which we call as electric vector fields. In addition to them, we introduce further $n_v+1$ vector fields~$A_{\mu\Lambda}$ called magnetic vector fields. They are summarized as $A_{\mu}^M=(A_\mu^\Lambda,A_{\mu\Lambda})^T$. Then, the generic gauging can be achieved by the so-called embedding tensor,
\begin{align}
    \Theta_{M}^{\ \alpha}=
    \begin{pmatrix} 
    \Theta_{\Lambda}^{\ \alpha} \\ \Theta^{\Lambda \alpha}
    \end{pmatrix},
    \label{ET}
\end{align}
and the modified covariant derivative,
\begin{align}
D_{\mu}\equiv&\;
\partial_{\mu}-A_{\mu}^{\Lambda}\Theta_{\Lambda}^{\ \alpha}T_{\alpha}-A_{\mu\Lambda}\Theta^{\Lambda \alpha}T_{\alpha}, \label{def_covder}
\end{align}
where all entries of Eq.~$\eqref{ET}$ are real and $\alpha$ denotes the isometries to be gauged. $T_{\alpha}$ is the generator of the isometries. 

The tensor~$\eqref{ET}$ must satisfy the following locality constraints,\footnote{For the case of non-Abelian gauging, some additional constraints are required.} 
\begin{align}
\Theta_{\Lambda}^{\ \alpha}\Theta^{\Lambda \beta}-\Theta_{\Lambda}^{\ \beta}\Theta^{\Lambda \alpha}=0. \label{cond_theta}
\end{align}
Also, to match the degree of freedom of vector fields, the auxiliary two-form fields~$B_{\mu\nu ,\alpha}$ and its (one-form) gauge symmetry should be introduced (see Refs.~\cite{Samtleben:2008pe,Trigiante:2016mnt} for review). 

Based on this technique, the relevant parts of the Lagrangian in this paper are given by 
\begin{align}
&\mathcal{L}_{vs}=\frac{1}{4}\mathcal{I}_{\Lambda \Sigma}\mathcal{H}_{\mu \nu }^{\Lambda }\mathcal{H}^{\Sigma \mu \nu }+\frac{i}{4}\mathcal{R}_{\Lambda \Sigma}\mathcal{H}_{\mu\nu}^{\Lambda }\tilde{\mathcal{H}}^{\Sigma \mu\nu}-\frac{i}{4}\Theta^{\Lambda \alpha}\tilde{B}_{\mu \nu ,\alpha}\left( F_{\Lambda}^{ \mu \nu}-\frac{1}{4}\Theta _{\Lambda}^{\ \beta}B^{\mu \nu }_{\beta}\right), \label{L}\\
&\mathcal{L}_{s}=h_{uv} D_{\mu} b^u D^{\mu} b^v+g_{i\Bar{j}} \partial_{\mu} z^i \partial^{\mu} \Bar{z}^{\Bar{j}}-V, \label{B}
\end{align}
where $\mathcal{R}_{\Lambda \Sigma}$ and $\mathcal{I}_{\Lambda \Sigma}$ are real and imaginary parts of gauge kinetic function $\mathcal{N}_{\Lambda \Sigma}$, respectively. 
The $\mathcal{H}$ is defined by the combination of the field-strength $F_{\mu \nu}^{\Lambda}=\partial_{\mu}A_{\nu}^{\Lambda}-\partial_{\nu}A_{\mu}^{\Lambda}$ and the auxiliary two-form~$B_{\mu\nu ,\alpha}$ as,
\begin{align}
\mathcal{H}_{\mu \nu}^{\Lambda} \equiv F_{\mu \nu}^{\Lambda}+\frac{1}{2}\Theta ^{\Lambda \alpha}B_{\mu\nu ,\alpha}, \label{def_H}
\end{align}
and $\tilde{\mathcal{H}}_{\mu\nu} \equiv -\frac{i}{2} \varepsilon_{\mu \nu \rho \sigma} \mathcal{H}^{\rho \sigma}$. The Lagrangian~$\eqref{B}$ contains kinetic terms for the scalars in vector and hypermultiplet and their potential, which is explicitly shown in Eq.~$\eqref{V}$. The hyper sector is gauged by Eq.~$\eqref{def_covder}$ with $T_{\alpha}b^u=\delta_{\alpha}^u$.

In this formalism, the model in Ref.~\cite{Ferrara:1995xi} corresponds to $n_v=1$ and a choice of 
\begin{align}
\Theta_{0}^{\ 2}=E\neq 0,\ \ \Theta^{1 1} =M\neq 0, \ \ {\rm{Others}}=0, \label{ET1}
\end{align}
which manifestly satisfies the constraint~$\eqref{cond_theta}$. In the following section, we call terms containing $E/M$ as electric/magnetic correction. 

From here, we construct the physical Lagrangian through two steps: integrating out two form fields and gauge away all extra fields. The details of integrating out two form and gauge fixing two form fields are given in Appendix~\ref{app:IntegratingOutTwoFormFields}.


\section{Model and mass spectra}\label{spectra}

In the model of Ref.~\cite{Ferrara:1995xi}, the two vector fields (graviphoton and a vector in vector multiplet) become massive, and therefore, there is no massless vector.\footnote{More generally, in the case with $n_v=1$, a complex scalar $(z^1)$ can have a mass only if the third derivative of the prepotential~$\left<f_{111}\right>$ is nontrivial. However, as shown in Ref.~\cite{Antoniadis:2018blk,Abe:2019vzi}, when $\left<f_{111}\right>\neq 0$, we always have the (partial breaking) vacuum where two vector fields become massive. Thus, we conclude that a massless vector cannot exist in the case with $n_v=1$.} To discuss the ALP-photon coupling, we need to introduce a massless vector which has nothing to do with the gauging, and it would be identified as photon. In the following, we consider the case with $n_v=2$ as a minimal setup, without changing the choice~$\eqref{ET1}$.


\subsection{Scalar sector}

In this subsection, we discuss the stationary conditions and the scalar masses. In this paper, we have two assumptions which simplify our calculation. First, we only consider the case of simple vacuum, that is, $\left<z^i\right>=0$. The second assumption is the form of prepotential being polynomial, and it takes expression as follows,
\begin{align}
f=c_0+c_iz^i+\frac{1}{2}c_{ij}z^iz^j+\frac{1}{6}c_{ijk}z^iz^jz^k,\label{prepotential}
\end{align}
where all of the coefficient are complex and symmetric under the interchange of their indices. Note that with the simple vacuum assumed, there is no need to have fourth or higher order terms for prepotential for our purpose.

Based on the quantities defined in Sec.~\ref{setup}, the scalar potential is given by
\begin{equation}
V=r_{MN}(g^{i\bar{j}} U^M_i  \bar{U}^N_{\bar{j}}-\bar{V}^MV^N),\label{V}
\end{equation}
where 
\begin{align} 
V^M\equiv e^{\mathcal{K}/2} \begin{pmatrix} 
X^{\Lambda} \\ F_{\Lambda}
\end{pmatrix}, \ \ U_i^M\equiv \left(\partial_i+\frac{1}{2}\partial_i\mathcal{K}\right)V^M, \ \ 
r_{MN}\equiv \frac{1}{(b^0)^2}\left(\sum_{\alpha=1}^3 \Theta_M^{\ \alpha}\Theta_N^{\ \alpha}\right).
\end{align}
The stationary conditions take forms, 
\begin{align} 
   & {\partial_{b^0} V=0}:\ \  r_{MN}(g^{i\bar{j}} U^M_i  \bar{U}^N_{\bar{j}}-\bar{V}^MV^N)=0,\label{Vb0} \\
    &{\partial_{i} V=0}:\ \  r_{MN}\bar{U}^M_{\bar{j}}\bar{U}^N_{\bar{k}}g^{j\bar{j}}g^{k\bar{k}}f_{ijk}=0.\label{Vzi} 
\end{align}
The analytic solution of these equations is in general difficult to obtain even in the simplified choice of the embedding tensor $\eqref{ET1}$, simple vacuum and the minimal form of prepotential~$\eqref{prepotential}$. Therefore, we concentrate on the case where at least one supersymmetry is conserved, which trivially satisfies Eq.~$\eqref{Vzi}$.

Furthermore, we assume that 
\begin{align} 
c_2=c_{12}=c_{112}=0,\ \ {\rm{Re}}c_{11}=0 \label{ass}
\end{align}
for simplicity,\footnote{This ansatz are used in Refs.~\cite{Itoyama:2006ef}.} which simplifies the derivatives of the K\"ahler potential at $\left<z^i\right>=0$, 
\begin{align}
\left<\partial_1\mathcal{K}\right>=-\frac{{\rm{Re}}c_1}{2{\rm{Re}}c_0}, \ \     \left<\partial_2\mathcal{K}\right>=0,
\end{align}
and diagonalizes the K\"ahler metric as
\begin{align}
\left<g_{i\bar{j}}\right>=\frac{1}{4({\rm{Re}}c_0)^2}\begin{pmatrix} 
({\rm{Re}}c_1)^2&0 \\ 0&-2{\rm{Re}}c_0{\rm{Re}}c_{22}
\end{pmatrix},
\end{align}
which implies that ${\rm{Re}}c_1\neq 0$, and ${\rm{Re}}c_{22}<0$ since ${\rm{Re}}c_0=\frac{1}{4}e^{-\left<\mathcal{K}\right>}>0$.

For $M\neq 0$, one can easily find that Eq.~$\eqref{Vb0}$ is rewritten as 
\begin{align}
{\rm{Im}}c_{11}\left({\rm{Re}}c_0{\rm{Im}}c_{11}-{\rm{Re}}c_1{\rm{Im}}c_1\right)=0.
\end{align}
Also, Eq.~$\eqref{Vzi}$ with $i=1$ gives 
\begin{align}
{\rm{Re}}c_1=\pm E/M,\ \ {\rm{and}}\ \ 2{\rm{Re}}c_0{\rm{Im}}c_{11}-{\rm{Re}}c_1{\rm{Im}}c_1=0,
\end{align}
under $c_{111}\neq 0$. Obviously, the consistent set of the solution is 
\begin{align}
{\rm{Im}}c_{1}={\rm{Im}}c_{11}=0, \ \ {\rm{Re}}c_1= E/M. \label{vaccond}
\end{align}
 Finally, Eq.~$\eqref{Vzi}$ with $i=2$ is trivially satisfied.

The second derivatives of the potential at this point are evaluated as
\begin{align}
\left<V_{1\bar{1}}\right>=\frac{{\rm{Re}}c_0|c_{111}|^2M^4}{(b^0)^2E^2}>0,\ \ \left<V_{2\bar{2}}\right>=-\frac{|c_{122}|^2M^2}{2(b^0)^2{\rm{Re}}c_{22}}>0, 
\end{align}
and the others vanish. Thus, we conclude that $\left<z^1\right>=\left<z^2\right>=0$ with Eqs.~$\eqref{ass}$ and $\eqref{vaccond}$ is at least local minimum of the potential. 

Expanding $z^i$ around the vacuum as $z^i\rightarrow \left<z^i\right>+z^i$ and taking into account the canonical normalization, 
\begin{align}
\mathcal{Z}^i\equiv \sqrt{\left<g_{ii}\right>}z^i,\ \ ({\rm{no\ sum\ for\ }}i),\label{c_z}
\end{align}
we obtain the scalar sector Lagrangian:
\begin{align}
\mathcal{L}=\sum_{i=1,2}\left(\partial_{\mu}\mathcal{Z}^i\partial^{\mu}\bar{\mathcal{Z}}^i-m_i^2|\mathcal{Z}^i|^2\right)+\cdots,
\end{align}
where 
\begin{align}
m_1^2=\frac{4({\rm{Re}}c_0)^3|c_{111}|^2M^6}{(b^0)^2E^4}, \ \ m_2^2=\frac{{\rm{Re}}c_0|c_{122}|^2M^2}{(b^0)^2({\rm{Re}}c_{22})^2},\label{m12}
\end{align}
and the ellipsis denotes higher order terms. In Sec.~\ref{VScoupling}, we discuss the couplings between $\mathcal{Z}^i$ and the vector fields.


\subsection{Vector sector}

Now we evaluate the vector masses. First, the kinetic terms of the vector fields are obtained in Appendix~\ref{quadcub}, and the result is the first term in Eq.~$\eqref{v_kin}$: 
\begin{align}
\mathcal{L}=&\frac{1}{4}\biggl[-{\rm{Re}}c_0\frac{M^2}{E^2}F_{1\mu\nu}F_1^{\mu\nu}-{\rm{Re}}c_0F^0_{\mu\nu}F^{0\mu\nu}+{\rm{Re}}c_{22}F^2_{\mu\nu}F^{2\mu\nu}\biggr]. 
\end{align}
Then, the canonical gauge fields~$\mathcal{A}_{\mu}$ are given by
\begin{align}
\sqrt{{\rm{Re}}c_0}\frac{M}{E}A_{1\mu}=\mathcal{A}_{1\mu}, \ \   \sqrt{{\rm{Re}}c_0}A_{\mu}^0=\mathcal{A}_{\mu}^0, \ \ \sqrt{|{\rm{Re}}c_{22}|}A_{\mu}^2=\mathcal{A}_{\mu}^2.  \label{c_v}
\end{align}

Also, from the kinetic terms for hyper sector in Eq.~$\eqref{B}$, we have
\begin{align}
h_{uv}D_{\mu}b^uD^{\mu}b^v=\frac{1}{2(b^0)^2}\biggl\{(\partial_{\mu}b^0)^2+(\partial_{\mu}b^3)^2+\left(\partial_{\mu}b^1-MA_{\mu 1}\right)^2+\left(\partial_{\mu}b^2-EA_{\mu }^0\right)^2\biggr\}.\label{stukelberg}
\end{align}
Then after fixing U(1) gauge symmetry by~$A_{\mu 1}\rightarrow A_{\mu 1}+\partial_{\mu}b^1/M$ and~$A_{\mu }^0\rightarrow A_{\mu }^0+\partial_{\mu}b^2/E$, the third and the fourth terms in Eq.~$\eqref{stukelberg}$ give the same masses for $\mathcal{A}_{\mu 1}$ and $\mathcal{A}_{\mu }^0$,
\begin{align} 
m^2_{\mathcal{A}}\equiv \frac{E^2}{(b^0)^2{\rm{Re}}c_0}.\label{m_A}
\end{align}  
The degeneration of their masses are the result of the remaining $\mathcal{N}=1$ supersymmetry. Indeed, they constitute a massive spin $3/2$ multiplets with massive gravitino and another fermion~\cite{Ferrara:1983gn}. Note that the gauge field $\mathcal{A}_{\mu}^2$ remains massless as expected, and we identify it as ``photon". As for the other two massive gauge fields, we assume those two massive gauge fields, $\mathcal{A}_{\mu}^0$ and $\mathcal{A}_{\mu 1}$, live in dark sector, and call them ``dark photons".


\section{ALP-photon Couplings}\label{VScoupling}

In this section, we focus on the ALP-photon couplings in the model given in Sec. \ref{spectra} and discuss the consequences from observations. From Eq.~\eqref{derR} with the normalization~$\eqref{c_z}$ and~$\eqref{c_v}$, and by omitting spacetime indices, we obtain
\begin{equation} \label{vs}
    \begin{split}
        \mathcal{L}
        =& g_1 \mathcal{Z}^1 \mathcal{F}_1 \tilde{\mathcal{F}}_1 
        + g_2 \mathcal{Z}^1 \mathcal{F}^0 \tilde{\mathcal{F}}^0 
        + g_3\mathcal{Z}^2 \mathcal{\mathcal{F}}^0 \tilde{\mathcal{F}}^2 
        + (g_4 \mathcal{Z}^1 + g_5 \mathcal{Z}^2) \mathcal{F}^2 \tilde{\mathcal{F}}^2 \\
        &+ i g_6 \mathcal{Z}^1 \mathcal{F}_1 \tilde{\mathcal{F}}^0 
        + i g_7 \mathcal{Z}^2 \mathcal{F}_1 \tilde{\mathcal{F}}^2 + {\rm{h.c.}},
    \end{split}
\end{equation}
where $\mathcal{F}$ are field-strengths defined by the canonical fields in Eq.~$\eqref{c_v}$. The coupling constants are given by
\begin{align}
    &g_1
    =\frac{1}{4M_{\rm{P}}} \left( 1 - ({\rm{Re}}c_0)^2 \ c_{111} \frac{M^3}{E^3} (M_{\rm{P}})^3 \right),
    \ \ 
    g_2
    =-g_1,\label{g1g2}
    \\
    &g_3
    =-\frac{\sqrt{2}}{4M_{\rm{P}}} \left( 1 + \frac{{\rm{Re}}c_0 \ c_{122}}{|{\rm{Re}}c_{22}|} \frac{M}{E} M_{\rm{P}} \right),
    \ \ 
    g_4
    =\frac{1}{4M_{\rm{P}}} \left( \frac{{\rm{Re}}c_0 \ c_{122}}{|{\rm{Re}}c_{22}|} \frac{M}{E} M_{\rm{P}} \right),\label{g3g4}
    \\
    &g_5
    =\frac{1}{4\sqrt{2}M_{\rm{P}}} \frac{\sqrt{{\rm{Re}}c_0}}{|{\rm{Re}}c_{22}|^{3/2}} c_{222},
    \ \ 
    g_6
    =\frac{1}{2M_{\rm{P}}} \left(({\rm{Re}}c_0)^2 \ c_{111} \frac{M^3}{E^3} (M_{\rm{P}})^3\right),
    \ \
    g_7
    =g_3,\label{g5g6}
\end{align}
where the reduced Planck mass $M_{\rm{P}}= 2.4\times 10^{18}$ GeV is recovered. Since we have $\mathcal{A}^2$ to be photon, the ALP-photon couplings are those with $g_4$ and $g_5$ coefficients.



Before going through the analysis of Eqs.~\eqref{vs} - \eqref{g5g6}, there are two points we wish to recall with an assumption $\langle b^0 \rangle \approx M_{\rm{P}}$\footnote{The stabilization of moduli fields is beyond the scale of this paper.}. First, the first supersymmetry breaking scale is given by Eq.~\eqref{m_A}, which is rewritten here but with the Planck scale, 
\begin{equation} \label{eq:susyScale}
    m_{\rm{SUSY}}\equiv \frac{E}{\sqrt{{\rm{Re}}c_0}} (M_{\rm{P}}).
\end{equation}
Next, the axion masses are given as follows with Planck mass given explicitly,
\begin{equation} \label{mpm12}
    \begin{split}
        \abs{m_1} &= \frac{2 {\rm{Re}}c_0\ \sqrt{{\rm{Re}}c_0}\ |c_{111}|\ M^3}{ E^2} (M_{\rm{P}})^4, \\ 
        \abs{m_2} &= \frac{\sqrt{{\rm{Re}}c_0}\ |c_{122}|\ M} { |{\rm{Re}}c_{22}|} (M_{\rm{P}})^2.
    \end{split}
\end{equation}
Then, we define relative mass scales for axions comparing to supersymmetry breaking scale. That is,
\begin{equation}
    \begin{split}
        \rho_1 
        &\equiv \frac{\abs{m_1}}{m_{\rm{SUSY}}} 
        = 2({\rm{Re}}c_0)^2\ |c_{111}|\ \frac{M^3}{E^3} (M_{\rm{P}})^3, \\ 
        \rho_2 
        &\equiv \frac{\abs{m_2}}{m_{\rm{SUSY}}} 
        = \frac{{\rm{Re}}c_0\ |c_{122}|}{|{\rm{Re}}c_{22}|} \frac{M}{E} M_{\rm{P}}.\label{rho12}
    \end{split}
\end{equation}

Now, we come back to the analysis of Eqs.~\eqref{vs} - \eqref{g5g6}. 

In Eqs.~\eqref{g1g2} - \eqref{g5g6}, we can find some corrections which depend on the gauge coupling~$E$. Also, from Eq.~\eqref{eq:susyScale}, we can directly relate $E$ to the supersymmetry breaking scale for fixed ${\rm{Re}}c_0$. Therefore, the corrections are the result of the first supersymmetry breaking. As the breaking scale becomes sufficiently low (small $E$), it enlarges some couplings in Eqs.~\eqref{g1g2} - \eqref{g5g6}. Besides, as we see that the corrections (terms depend on $E$ or $M$) in Eqs.~\eqref{g1g2} - \eqref{g5g6} are related to the axion masses in Eq.~\eqref{mpm12}, and interestingly, those terms are explicitly proportional to $\rho_1$ and $\rho_2$ in Eq.~\eqref{rho12}. That, then, not only suggests axion masses are related to couplings but also implies that the first supersymmetry breaking scale is influencing the axion phenomenology in this model.

Since $\mathcal{A}^0$ and $\mathcal{A}_1$ stay in dark sector while $\mathcal{A}^2$ is in normal sector, we have a classification for axions from Eq.~\eqref{vs}; that is, except for ALP-photon couplings ($g_4$ and $g_5$ terms), $\mathcal{Z}^1$ is responsible to interact within dark sector ($g_1$, $g_2$ and $g_6$ terms), while $\mathcal{Z}^2$ interacts between normal and dark sectors ($g_3$ and $g_7$ terms). 

Having two different types of complex axions brings us a difficulty of how to compare to observations, which most of the works are devoted in only one kind of axions. Therefore, to constrain our model with axion observations, we, for simplicity, consider about its asymptotic behavior which makes the Lagrangian \eqref{vs} contains either $\mathcal{Z}^1$ or $\mathcal{Z}^2$. Here we demonstrate three ways of completing the asymptotic analyses: (i) $\rho_1,\rho_2 \ll 1$, (ii) $\rho_2\ll1\ll \rho_1 $ and (iii) $\rho_1\ll1\ll \rho_2$. Also, as we will see later
, a mild assumption on the parameters $c_{ijk}$ in prepotential leads to the Lagrangian of an effectively single axion field, which simplifies the situation.


Note that in the following discussion, we use the axion observational constraint from Ref.~\cite{Anastassopoulos:2017ftl}. That is, axion-photon coupling $g_{a\gamma\gamma}$  is bounded with given axion mass $m$ range as $g_{a\gamma\gamma} \lesssim 0.66 \times 10^{-10} \ \rm{GeV}^{-1}$ when $m \lesssim 0.02\ \rm{eV}$, and for simplicity we implement their product as our constraint, which is looser. That is,
\begin{equation} \label{eq:constraint}
    mg_{a\gamma\gamma} \lesssim 1.32 \times 10^{-21}.
\end{equation}

\subsection{$\rho_1,\rho_2 \ll 1$} \label{sec:stableAxion}


To get started with, we assume both axion masses are relatively small comparing to the first breaking scale. That means $\rho_1,\rho_2\ll 1$, and for simplicity, we assume that it is caused by having $c_{111},c_{122} \rightarrow 0$\footnote{This assumption is not necessary because couplings or its magnetic corrections are proportional to $\rho_1$ and $\rho_2$. However, this assumption simplifies the analysis without deterring the final result.}. Then, the effective Lagrangian becomes,
\begin{equation}\label{eff_1}
    \begin{split}
        \mathcal{L}_{eff}
        =& \ g_1 \mathcal{Z}^1 \mathcal{F}_1 \tilde{\mathcal{F}}_1 
        + g_2 \mathcal{Z}^1 \mathcal{F}^0 \tilde{\mathcal{F}}^0 
        + g_3\mathcal{Z}^2 \mathcal{\mathcal{F}}^0 \tilde{\mathcal{F}}^2 
        + g_5 \mathcal{Z}^2 \mathcal{F}^2 \tilde{\mathcal{F}}^2 
        + i g_7 \mathcal{Z}^2 \mathcal{F}_1 \tilde{\mathcal{F}}^2 + {\rm{h.c.}} \\
        \equiv&\ ig_{a \gamma \gamma} \ a \mathcal{F} \tilde{\mathcal{F}}
        + i g_{a \gamma \rm{H}} \ a \mathcal{F}_{\rm{H}} \tilde{\mathcal{F}}
        +\cdots,
    \end{split}
\end{equation}
where the ellipsis denotes terms not containing axion defined in Eq.~\eqref{eq:effAxionCoup} and the couplings become,
\begin{align} \label{4.1:g'1g'2}
   & g_1\sim\frac{1}{4M_{\rm{P}}}, \ \ g_2=-g_1,\\
  &  g_3\sim -\frac{\sqrt{2}}{4M_{\rm{P}}},
    \ \ g_5=\frac{1}{4\sqrt{2}M_{\rm{P}}}\frac{\sqrt{{\rm{Re}}c_0}}{|{\rm{Re}}c_{22}|^{3/2}}c_{222},
    \ \ g_7=g_3. \label{g'3g'5g'7}
\end{align}
Here by assuming $c_{222}$ to be real for simplicity, and then, the effective fields and associated couplings in the second equality of Eq.~\eqref{eff_1} are given as,
\begin{equation} \label{eq:effAxionCoup}
    \begin{split}
        &a \equiv \rm{Im}\mathcal{Z}^2, 
        \ \ \mathcal{F} \equiv \mathcal{F}^2, 
        \ \ \mathcal{F}_{\rm{H}} \equiv \mathcal{F}^0, \\
        &g_{a \gamma \gamma} \equiv 2g_5,
        \ \ g_{a \gamma \rm{H}} \equiv 2g_3.
    \end{split}
\end{equation}
where $a$ is axion field, $\mathcal{F}$ is photon field-strength, and $\mathcal{F}_{\rm{H}}$ is hidden photon field-strength.

Here, it is then clear that $\mathcal{Z}^1$ sink into the dark sector entirely. The reason is the assumption of $\rho_2 \ll 1$. Also, note that the only coupling constant which depends on the form of the prepotential is that associated with the axionic coupling, $g_{a \gamma \gamma }$, while the rest are just small constants. 



Here we show the possibility of constraining the theory by applying the constraint \eqref{eq:constraint} to the product of $g_{a \gamma \gamma}$ in Eq.~\eqref{eq:effAxionCoup} and $m_2$ in Eq.~\eqref{mpm12} (since $a={\rm{Im}}\mathcal{Z}^2$ is the effective axion), and we get,
\begin{equation} \label{constraint4.1}
    \begin{split}
        |m_2 g_{a\gamma\gamma}| 
        &= \frac{M_{\rm{P}}}{2\sqrt{2}} \Big(\frac{{\rm{Re}}c_0}{|{\rm{Re}}c_{22}|^{5/2}} |c_{122}| |c_{222}| \Big) M 
        \lesssim 1.32 \times 10^{-21},
    \end{split}
\end{equation}
which is equivalent to 
\begin{align}
c_{\rm{pre}}M   \lesssim 3.73 \times 10^{-21},
\end{align}
where we defined a dimensionless combination of the parameters in prepotential as $c_{\rm{pre}}={\rm{Re}}c_0|c_{122}| |c_{222}|M_{\rm{P}}/|{\rm{Re}}c_{22}|^{5/2}$. Then, we obtained the upper bound for the product of parameters in prepotential, $c_{\rm{pre}}$, and gauge coupling, $M$. This observation would be important when we embed the model into string models since those parameters are determined by the flux and geometric quantities in that case (see Ref.~\cite{Cicoli:2012sz} for example).
Interestingly, this upper bound is independent of another gauge parameter $E$, and thus, it is independent of the first breaking scale given in Eq.~\eqref{eq:susyScale} with fixed ${\rm{Re}}c_0$.

\subsection{$\rho_1 \gg 1 \gg \rho_2 $} \label{sec:decouplingz1}
Next, let us consider the limit $\mathcal{Z}^1$ decouples. This can be achieved by $\rho_1\gg 1\gg \rho_2$, or for simplicity,  $|c_{111}|\gg |c_{122}|$.  Then the effective Lagrangian contains the following terms,
\begin{align}
 \nonumber \mathcal{L}_{eff}=&\ g_3\mathcal{Z}^2\mathcal{\mathcal{F}}^0\tilde{\mathcal{F}}^2+g_5\mathcal{Z}^2\mathcal{\mathcal{F}}^2\tilde{\mathcal{F}}^2+ig_7\mathcal{Z}^2\mathcal{F}_1\tilde{\mathcal{F}}^2+{\rm{h.c.}},\\
 \equiv &\ ig_{a\gamma\gamma}a\mathcal{\mathcal{F}}\tilde{\mathcal{F}}+ig_{a\gamma H}a\mathcal{\mathcal{F}}\tilde{\mathcal{F}}_{\rm{H}} +\cdots, \label{vs_EFT2}
\end{align}
where the definitions of the fields and the couplings in the second line are the same with Eq.~\eqref{eq:effAxionCoup}. Therefore, the terms containing axion is exactly same with the example we discussed in Sec.~\ref{sec:stableAxion}. Then the constraints~\eqref{constraint4.1} obtained in the previous section can also be applied in this case. Note that the similarities are only in ALP-photon couplings. The other parts between these two cases are totally different. That is, to fully determine which case we fall into, we need to discuss the other terms in the Lagrangians.

\subsection{$\rho_2 \gg 1 \gg \rho_1 $}
\label{sec:decouplingz2}
Similar to Sec.~\ref{sec:decouplingz1}, by assuming $\rho_2 \gg 1 \gg \rho_1 $, or for simplicity $|c_{122}|\gg |c_{111}|$, we get decoupling limit of $\mathcal{Z}^2$. Then the effective Lagrangian is obtained as,
\begin{equation} \label{vs_EFT}
    \begin{split}
        \mathcal{L}_{eff}
        =&\ g_1 \mathcal{Z}^1 \mathcal{F}_1 \tilde{\mathcal{F}}_1 
        + g_2 \mathcal{Z}^1 \mathcal{F}^0 \tilde{\mathcal{F}}^0 
        + g_4 \mathcal{Z}^1 \mathcal{F}^2 \tilde{\mathcal{F}}^2 
        + i g_6 \mathcal{Z}^1 \mathcal{F}_1 \tilde{\mathcal{F}}^0 
        + {\rm{h.c.}},\\
        \equiv&\ 
         ig_{a\gamma\gamma} \ a \mathcal{F} \tilde{\mathcal{F}}
        + ig_{a\rm{H}\rm{H}} \ a \mathcal{F}_{\rm{H}} \tilde{\mathcal{F}}_{\rm{H}} 
        + ig_{a\rm{H}'\rm{H}'} \ a \mathcal{F}_{\rm{H}'} \tilde{\mathcal{F}}_{\rm{H}'}
        + \cdots
    \end{split}
\end{equation}
where the ellipsis denotes terms without axion which is defined in Eq.~\eqref{eq:4.2_effAxion}, and coupling constants are,
\begin{align} 
    &g_1 \sim \frac{1}{4M_{\rm{P}}}, \ \ g_2=-g_1, \\ \label{4.2:g'1g'2}
    &g_4 = \frac{1}{4}\frac{M}{E}\frac{{\rm{Re}}c_0}{|{\rm{Re}}c_{22}|}c_{122},
    \ \ g_6 = \frac{(M_{\rm{P}})^2}{2}\frac{M^3}{E^3}({\rm{Re}}c_0)^2c_{111}.
\end{align}
Here by assuming $c_{122}$ to be real for simplicity, and then, the effective fields and associated couplings in the second equality of Eq.~\eqref{vs_EFT} are given as,
\begin{equation} \label{eq:4.2_effAxion}
    \begin{split}
        &a \equiv \rm{Im}\mathcal{Z}^1, 
        \ \ \mathcal{F} \equiv \mathcal{F}^2, 
        \ \ \mathcal{F}_{\rm{H}} \equiv \mathcal{F}^0,
        \ \ \mathcal{F}_{\rm{H}'} \equiv \mathcal{F}_1, \\
        &g_{a\gamma\gamma} \equiv 2g_4,
        \ \ g_{a\rm{H}\rm{H}} \equiv 2g_2, 
        \ \ g_{a\rm{H}'\rm{H}'} \equiv 2g_1,
    \end{split}
\end{equation}
where $a$ is axion field, $\mathcal{F}$ is photon field-strength, and $\mathcal{F}_{\rm{H}^{(')}}$ are hidden photon field-strengths.

 Similar to what we had in Sec. \ref{sec:stableAxion}, only $g_{a\gamma\gamma}$ is the concerning term, the ALP-photon coupling. Then, by applying Eq.~\eqref{eq:constraint} to the product of $m_1$ in Eq.~\eqref{mpm12} and $g_{a\gamma\gamma}$ in Eq.~\eqref{eq:4.2_effAxion}, we get,
\begin{equation} \label{eq:constraint4.3}
    \begin{split}
        |m_1 g_{a \gamma\gamma}|
        &= (M_{\rm{P}})^4 \Big( \frac{({\rm{Re}}c_0 )^{5/2}}{|{\rm{Re}}c_{22}|} |c_{111}| |c_{122}| \Big) \frac{M^4}{E^3}
        \lesssim 1.32 \times 10^{-21},\\
    \end{split}
\end{equation}
or equivalently,
\begin{align} \label{eq:constraint4.3v2}
c_{\rm{pre}}\frac{M^4}{E^3}   \lesssim 1.32 \times 10^{-21}, 
\end{align}
where $c_{\rm{pre}}\equiv ({\rm{Re}}c_0 )^{5/2}|c_{111}| |c_{122}|(M_{\rm{P}})^4/|{\rm{Re}}c_{22}| $.
In the same way with Eq.~\eqref{constraint4.1}, this gives an upper bound for the product of some parameters of prepotential and gauge couplings. However, the left-hand side of Eq.~\eqref{eq:constraint4.3v2} depends also on $E$ in this case, which is related to the supersymmetry breaking scale given by Eq.~\eqref{eq:susyScale} and differ from the previous cases. Then we can rewrite Eq.~\eqref{eq:constraint4.3} in terms of the first breaking scale as,
\begin{equation} \label{eq:msusy4.3}
    {m_{\rm{SUSY}}}
    \gtrsim M_{\rm{P}}c'_{\rm{pre}} M^{4/3}    \times 10^{7},
\end{equation}
where $c'_{\rm{pre}}\equiv({\rm{Re}}c_0|c_{111}||c_{122}|/|{\rm{Re}c_{22}}|(M_{\rm{P}})^4)^{1/3}$ is again a dimensionless collective of the parameters inside the prepotential. 
Then we see that the lower bound of the first supersymmetry breaking scale can be derived by axion observations.
The value of~$c'_{\rm{pre}} M^{4/3}$ is not determined in our setup, but it should be fixed if we consider the compactification of higher dimensional supergravity concretely and we can estimate how large the first breaking scale should be.  

\section{Summary} \label{summary}

In this paper, we demonstrated the possibility of restricting several parameters in extended supergravity with observerations; to be specific, we took advantage of axion observations to constrain gauged $\mathcal{N}=2$ supergravity.
In supersymmetric theory, couplings with neutral scalars and vector fields in general exist through the gauge kinetic function. In particular, with the help of ALP-photon observations the coupling between neutral scalars and photon and scalar masses, in principle, can be restricted, if we regard such scalars as ``axion".  Then, we obtained the constraints on the parameters of the theory which could be compared to string compactification. Also, note that the gauge kinetic function and the scalar potential are related in $\mathcal{N}=2$ theory, which leads to non-trivial relation between ALP-photon couplings and ALP masses.

Besides that, as mentioned in Introduction, restrictions on the breaking scale of extended supersymmetry are important from the phenomenological viewpoints, but were rarely discussed. In this paper, we focus on the partial breaking model~\cite{Ferrara:1995xi}, and show that the breaking scale of $\mathcal{N}=2\rightarrow\mathcal{N}=1$ can be also constrained by the ALP-photon constraints.

At the same time, the model we discussed contains two types of axions with different set of couplings. One of them ($\mathcal{Z}^1$) is responsible to interactions within either dark or normal sector themselves, while the other one ($\mathcal{Z}^2$) interacts with both sectors at once. These features seem interesting in terms of the several applications of axion-photon-hidden photon, and axion-double hidden photon couplings, which are discussed in e.g.,~
\cite{Kaneta:2016wvf,Kaneta:2017wfh,Alvarez:2017eoe,Kitajima:2017peg,Daido:2018dmu,Choi:2018dqr,Choi:2018mvk,Choi:2019jwx}.

For practical reasons, we considered three different effective theory of generating only single axion just for simplicity. In Sec.~\ref{sec:stableAxion}, by asking both types of axions being relatively stable (light) and with mild assumption on the parameter in prepotential, we find that  ${\rm{Im}}\mathcal{Z}^2$ is responsible for ALP-photon coupling. Then we obtain a bound for the parameters in prepotential and gauge coupling as Eq.~\eqref{eq:constraint4.3v2}.
In Sec.~\ref{sec:decouplingz1} where we consider the limit $\mathcal{Z}^1$ decouples, a similar result with that of Sec.~\ref{sec:stableAxion} is obtained, when we focus only on the axion interactions. Finally, in Sec.~\ref{sec:decouplingz2}, we consider the case $\mathcal{Z}^1$ decouples and find the bound~\eqref{eq:constraint4.3v2} for the parameters. Indeed, this can be rewritten as Eq.~\eqref{eq:msusy4.3}, which gives a suggestive lower bound for the supersymmetry breaking scale.

There are several remarks : 
(i) As mentioned before, the ``axion" in this paper is different from the usual one~\cite{Peccei:1977hh,Peccei:1977ur} or other ALPs~\cite{Marsh:2015xka}  which originate from global symmetry breaking as Nambu-Goldstone boson. The scalar fields we discussed are not protected by the shift symmetry if we consider other sectors coupling with it. 
Thus, the other terms besides ALP-photon couplings have different properties compared to the ordinary ones. (ii) Here we focused on the model of Ref.~\cite{Ferrara:1995xi} where $\mathcal{N}=1$ supersymmetry remains unbroken. It is interesting to investigate how the breaking from $\mathcal{N}=1$ to $\mathcal{N}=0$ affects on the result. (iii) Our approach can be applied to other extended supergravity models. 
It is also interesting to study other models and see direct restrictions on flux and geometrical parameters defined in compactification.

\subsection*{Acknowledgements}
S.~A. is supported in part by a Waseda University Grant for Special Research Projects (Project
number: 2019E-059). 

\appendix


\section{Integrating out two form fields}\label{app:IntegratingOutTwoFormFields}

Here we show the resultant Lagrangian after the integration of the two form-fields. For the general arguments, we refer Refs.~\cite{deWit:2002vt,deWit:2005ub,Samtleben:2008pe,Trigiante:2016mnt,DallAgata:2003sjo,DAuria:2004yjt,Andrianopoli:2011zj}. In the following discussion, we will suppress spacetime indices.     

The relevant parts are given in Eq.~$\eqref{L}$. In the case with $n_v=2$, or $\Lambda=0,1,2$, and with our choice of the embedding tensor~$\eqref{ET1}$, they are reduced to
\begin{align}
&\mathcal{L}=\frac{1}{4}\mathcal{I}_{\Lambda \Sigma}\mathcal{H}^{\Lambda }\mathcal{H}^{\Sigma  }+\frac{i}{4}\mathcal{R}_{\Lambda \Sigma}\mathcal{H}^{\Lambda }\tilde{\mathcal{H}}^{\Sigma }-\frac{i}{4}M\tilde{B}_{ 1}\left( F_{1}-\frac{1}{4}EB_{1}\right), \label{L2}
\end{align}
where 
\begin{align}
\mathcal{H}^{\Lambda}=\begin{pmatrix} 
F^0 \\ F^1+\frac{1}{2}MB_1\\F^2
\end{pmatrix}.
\end{align}
Note that the Lagrangian~$\eqref{L}$ is invariant up to total derivative under
\begin{align}
&\delta B_{\mu\nu,1}=\partial_{\mu}\Xi_{\nu,1}-\partial_{\nu}\Xi_{\mu,1},\\
&\delta A_{\mu}^{\Lambda}=\partial_{\mu}\lambda^{\Lambda}-\frac{1}{2}M\Xi_{\mu,1},\\
&\delta A_{\mu\Lambda}=\partial_{\mu}\lambda_{\Lambda},
\end{align}
where $\lambda^{\Lambda}$ and $\Xi_{1}$ are the parameters of the ordinary zero-form and the one-form gauge symmetries, respectively. Using the gauge symmetry, we can eliminate $F^1$ by
\begin{align}
B_1\rightarrow B_1-\frac{2}{M}F^1.    \label{B1gague}
\end{align}
Then it is straightforward to solve the E.O.M of $B_1$ which gives 
\begin{align}
B_1=\frac{2}{M}\left(\mathcal{I}_{11}+r^2\mathcal{I}_{11}^{-1}\right)^{-1}\left(\tilde{J}-ir\mathcal{I}_{11}^{-1}J\right),\label{solB}
\end{align}
where 
\begin{align}
r\equiv \mathcal{R}_{11}+\frac{E}{M}, \ \ J\equiv -\mathcal{I}_{1U}\tilde{F}^U-i\mathcal{R}_{1U}F^U+iF_1.
\end{align}
The index $U$ runs only $0$ and $2$. 
Substituting the solution~$\eqref{solB}$ into the Lagrangian, we obtain
\begin{align} 
       \nonumber  \mathcal{L}=
        &\frac{1}{4}
        \big( 
        \hat{\mathcal{I}}^{11} F_1 F_1  +\hat{\mathcal{I}}_{UV} F^U F^V
        +2 {\hat{\mathcal{I}}^{1}}_{\ U} F_1 F^U 
        \big)\\
        &+\frac{i}{4}
        \big( 
        \hat{\mathcal{R}}^{11} F_1 \tilde{F}_1  +\hat{\mathcal{R}}_{UV} F^U \tilde{F}^V
        +2 {\hat{\mathcal{R}}^{1}}_{\ U} F_1 \tilde{F}^U 
        \big),\label{onshell}
\end{align}
where
\begin{align} 
    &\hat{\mathcal{I}}^{11}
    =\left(\mathcal{I}_{11}+r^2\mathcal{I}_{11}^{-1}\right)^{-1},\\
    &\hat{\mathcal{I}}_{UV}
    = \mathcal{I}_{UV}+\hat{\mathcal{I}}^{11}\big( \mathcal{R}_{1U} \mathcal{R}_{1V} - \mathcal{I}_{1U} \mathcal{I}_{1V} - 2 r\mathcal{I}_{11}^{-1} \mathcal{R}_{1(U} \mathcal{I}_{V)1} \big),\\
    &{\hat{\mathcal{I}}^{I}}_{\ U}
    = \hat{\mathcal{I}}^{11} \big( -\mathcal{R}_{1U} +r \mathcal{I}_{11}^{-1}\mathcal{I}_{1U} \big),\\
    &\hat{\mathcal{R}}^{11}
    =  -\hat{\mathcal{I}}^{11} r\mathcal{I}_{11}^{-1},\\
    &\hat{\mathcal{R}}_{UV}
    = \mathcal{R}_{UV}
    +\hat{\mathcal{I}}^{11} \big(r\mathcal{I}_{11}^{-1} (-\mathcal{R}_{1U} \mathcal{R}_{1V} + \mathcal{I}_{1U} \mathcal{I}_{1V}) 
    -2 \mathcal{R}_{1(U} \mathcal{I}_{V)1} \big),\\
    &{\hat{\mathcal{R}}^{I}}_{\ U}
    = \hat{\mathcal{I}}^{11}\big( \mathcal{I}_{1U} + r\mathcal{I}_{11}^{-1}\mathcal{R}_{1U} \big).
\end{align}
Note that the degree of freedom of vector fields are $3$, i.e., $A^0, A_1$ and $A^2$, even though we introduced a magnetic vector $A_1$ since $A_{0,2}$ are absent from the beginning and $A^1$ is gauged away in Eq.~$\eqref{B1gague}$.


\section{Couplings at the vacuum}\label{quadcub}

Here we show the result of relevant couplings evaluated at the vacuum. Expanding the complex scalars around their vacuum expectation values, the quadratic and the cubic terms of Eq.~$\eqref{onshell}$ are given by
\begin{align}
\nonumber \mathcal{L}_{\rm{quad}}=&\frac{1}{4}\biggl[\left< \hat{\mathcal{I}}^{11}\right>F_1F_1+\left< \hat{\mathcal{I}}_{UV}\right>F^UF^V+2\left< \hat{\mathcal{I}}^1_{\ U}\right>F_1F^U\biggr]\\
&+\frac{i}{4}\biggl[\left< \hat{\mathcal{R}}^{11}\right>F_1\tilde{F}_1+\left< \hat{\mathcal{R}}_{UV}\right>F^U\tilde{F}^V+2\left< \hat{\mathcal{R}}^1_{\ U}\right>F_1\tilde{F}^U\biggr],\label{v_kin}\\
\nonumber \mathcal{L}_{\rm{cub}}=&\frac{1}{4}\biggl[\left< \partial_m\hat{\mathcal{I}}^{11}\right>z^mF_1F_1+\left< \partial_m\hat{\mathcal{I}}_{UV}\right>z^mF^UF^V+2\left< \partial_m\hat{\mathcal{I}}^1_{\ U}\right>z^mF_1F^U\biggr]\\
&+\frac{i}{4}\biggl[\left<\partial_m \hat{\mathcal{R}}^{11}\right>z^mF_1\tilde{F}_1+\left<\partial_m \hat{\mathcal{R}}_{UV}\right>z^mF^U\tilde{F}^V+2\left< \partial_m\hat{\mathcal{R}}^1_{\ U}\right>z^mF_1\tilde{F}^U\biggr]+{\rm{h.c.}},
\end{align}
where 
\begin{align}
\left< \hat{\mathcal{I}}^{11}\right>=-{\rm{Re}}c_0\frac{M^2}{E^2}, \ \ \left< \hat{\mathcal{I}}_{UV}\right>=\begin{pmatrix} 
-{\rm{Re}}c_0& 0 \\ 0& {\rm{Re}}c_{22} 
\end{pmatrix}
,\ \ \left< \hat{\mathcal{I}}^1_{\ U}\right>=\begin{pmatrix} 
 0 \\ 0
\end{pmatrix}
\\
\left< \hat{\mathcal{R}}^{11}\right>=0, \ \ \left< \hat{\mathcal{R}}_{UV}\right>=\begin{pmatrix} 
2{\rm{Im}}c_0& 0 \\ 0& {\rm{Im}}c_{22} 
\end{pmatrix}
,\ \ \left< \hat{\mathcal{R}}^1_{\ U}\right>=\begin{pmatrix} 
\pm  {\rm{Re}}c_0\frac{M^2}{E^2} \\ 0
\end{pmatrix}
\end{align}
and 
\begin{align}
\nonumber &\left< \partial_m\hat{\mathcal{I}}^{11}\right>=\left\{-\frac{M}{2E}\left(1+\frac{M^3}{E^3}({\rm{Re}}c_0)^2c_{111}\right),0\right\},\\
\nonumber &\left< \partial_m\hat{\mathcal{I}}_{UV}\right>=\frac{1}{2}\left\{\begin{pmatrix} 
\frac{E}{M}+\frac{M^2}{E^2}({\rm{Re}}c_0)^2c_{111}& 0 \\ 0& c_{122} 
\end{pmatrix}, \begin{pmatrix} 
0&-{\rm{Re}}c_{22}-\frac{M}{E}{\rm{Re}}c_0c_{122} \\ -{\rm{Re}}c_{22}-\frac{M}{E}{\rm{Re}}c_0c_{122} &c_{222}
\end{pmatrix}\right\},\\
&\left< \partial_m\hat{\mathcal{I}}^1_{\ U}\right>=\frac{i}{2}\left\{\begin{pmatrix} 
\frac{M^3}{E^3}({\rm{Re}}c_0)^2c_{111}\\  0 
\end{pmatrix}, \begin{pmatrix} 
0 \\ -\frac{M}{E}{\rm{Re}}c_{22}-\frac{M^2}{E^2}{\rm{Re}}c_0c_{122}  
\end{pmatrix}\right\}.\\
\nonumber &\left< \partial_m\hat{\mathcal{R}}^{11}\right>=\left\{-\frac{iM}{2E}\left(1-\frac{M^3}{E^3}({\rm{Re}}c_0)^2c_{111}\right),0\right\},\\
\nonumber &\left< \partial_m\hat{\mathcal{R}}_{UV}\right>=\frac{i}{2}\left\{\begin{pmatrix} 
\frac{E}{M}-\frac{M^2}{E^2}({\rm{Re}}c_0)^2c_{111}& 0 \\ 0& -c_{122} 
\end{pmatrix}, \begin{pmatrix} 
0& -{\rm{Re}}c_{22}+\frac{M}{E}{\rm{Re}}c_0c_{122} \\ -{\rm{Re}}c_{22}+\frac{M}{E}{\rm{Re}}c_0c_{122}& -c_{222} 
\end{pmatrix}\right\},\\
&\left< \partial_m\hat{\mathcal{R}}^1_{\ U}\right>=\frac{1}{2}\left\{\begin{pmatrix} 
\frac{M^3}{E^3}({\rm{Re}}c_0)^2c_{111} \\  0 
\end{pmatrix}, \begin{pmatrix} 
0 \\ \frac{M}{E}{\rm{Re}}c_{22}-\frac{M^2}{E^2}{\rm{Re}}c_0c_{122} 
\end{pmatrix}\right\}.\label{derR}
\end{align}
Here we used a notation $\partial_m  \mathcal{I}=\{\partial_1 \mathcal{I},\partial_2 \mathcal{I}\}$ and so on.


\end{document}